\documentstyle[twocolumn,prl,aps]{revtex}

\input{psfig}
\begin{document}
\preprint{}
\draft
\title{Fractional pumping of energy into a ballistic quantum ring.}
\author{L. Gorelik,$^{(1,2)}$ S. Kulinich,$^{(1,2)}$
Yu. Galperin,$^{(3)}$ R. I. Shekhter,$^{(1)}$ 
and M. Jonson$^{(1)}$ } 
\address{
     $^{(1)}$Department of Applied Physics,
     Chalmers University of Technology and G{\"o}teborg University,
     S-412 96 G\"oteborg, Sweden    \\
$^{(2)}$ B. Verkin Institute for Low Temperature Physics and Engineering, 310164 Kharkov, Ukraine\\
$^{(3)}$Department of Physics, University of Oslo, P.O. Box 1048 Blindern
0316 Oslo, Norway, \\ and A. F. Ioffe Institute, 194021 St.
Petersburg, Russia 
\\ }
\maketitle
\begin{abstract}

We consider the energy stored in a one-dimensional ballistic ring with
a barrier subject to a linearly time-dependent magnetic flux. 
An {\em exact analytic solution} for the quantum dynamics of electrons
in the ring is found for the case when the electro-motive force ${\cal E}$
is much smaller than the level spacing, $\Delta$.
Electron states exponentially localized in energy are found for irrational
values of the ratio $A\equiv\Delta/2e{\cal E}$. Relaxation limits 
the dynamic evolution and localization does 
not develop if $A$ is sufficiently close to a rational number. As a result
the accumulated energy becomes a regular function of $A$
containing a set of sharp peaks at rational values (fractional
pumping).
The shape of the peaks and the distances between them are 
governed by the interplay between the strength of backscattering
and the relaxation rate.
\end{abstract}
\pacs{73.40.-c, 73.2.Dx}

\narrowtext
Physical properties of mesoscopic systems are governed by quantum
interference. Several phenonema of such a nature have been discussed for
systems close to equilibrium. Persistent currents in
multiply-connected systems\cite{imry} as well as universal fluctuations of
the conductance are important examples\cite{alt}. 
Coherent dynamics remains crucially important in situations far
from equilibrium provided the energy associated with the 
phase breaking rate is less than the characteristic rate of redistributing
electrons in energy space. 
Consequently, one can expect
pronounced mesoscopic behavior even in strongly biased mesoscopic devices,
where the dynamics can be effectively tuned by external electric or magnetic
fields.

In this paper, we consider an example of such a system, namely a
single-channel mesoscopic ring subjected to a nonstationary
perpendicular magnetic field, linearly dependent on time. We
concentrate on the energy accumulation in such a system. To
investigate the role of interference, we take into account
electron backscattering  from a single potential barrier, embedded in
the ring. Tuning the transmission through the barrier by gate potentials
 one can influence
the interference pattern and in this way significantly change the
dynamics.

Impure onducting rings 
have been extensively discussed in connection with
energy dissipation in mesoscopic metallic systems. Gefen and
Thouless\cite{tg,gt} have suggested that disorder leads to a
localization of electrons {\em in energy space}. Consequently, a
time-dependent magnetic flux, $\Phi$,  through the ring
could induce a (dc)
circular current only in the presence of phase breaking
processes. This work was continued in Ref.~\onlinecite{g1} by a numerical 
analysis
of the role of dissipation. A numerical analysis of localization
in a dissipation-free impure ring was performed in Ref.~\onlinecite{blatter}.

The vanishing of the flux-induced current in an
impure dissipation-free ring\cite{tg,gt} is drastically different 
from what happens in a
perfect ballistic ring, where a non-equilibrium flux
driven current diverges when the dissipation goes to zero. This makes 
the crucial importance of the backscattering strength clear: By
tuning the height of the barrier one can cross over from one regime to
the other, and in this way control the pumping of energy
into the system. This issue, not addressed in previous
work, is the subject of the present paper. 

We will show that the scenario of the cross over is as
follows. Consider the conductance of the ring,
$G$, defined as the ratio between the circulating current and the
    electro-motive force ${\cal E} =-(1/c){\dot \Phi}$ induced in a ring
of radius $r_0$ by a magnetic field linearly dependent on
time. If scattering is strong, $G \propto {\cal E}^{-2}$. As the
scattering strength decreases, a set of peaks in the 
function $G=G({\cal E})$ appears. The peaks correspond to
rational values $p/q$ of the quantity $A\equiv\Delta/2e{\cal E}$, where
$\Delta=\hbar^2 N_F/mr_0^2$. Here 
$N_F$ is the number of filled electron states while $m$ is the effective mass.
The shape of the peaks and the distances
between them are 
governed by an interplay between the  height $V$ of the potential
barrier and the relaxation rate, $\nu$, the maximum value of $q$ being
determined by 
the condition $\tau^q /q \simeq \hbar\nu /e{\cal E}$. 
Here $\tau \equiv \exp(-{\cal
E}_c/{\cal E})$ is the effective amplitude of Zener tunneling through
the energy gaps in the electron spectrum, ${\cal E}_c =
\pi^2V^2/2\Delta e$. The peak structure
near maximum can be described by the interpolation formula
\begin{equation}\label{1}
g= \tau^{2q} \frac{\hbar\nu \Delta}{(\hbar\nu)^2+(e{\cal
E}q)^2\varepsilon^2} + \eta q^2 \frac{\hbar\nu \Delta }{(e{\cal E})^2} 
,
\ \varepsilon = A-\frac{p}{q} 
\end{equation} 
Here $g\equiv G/G_0, \ G_0 = e^2/h$, while $\eta(\varepsilon)$ is
a smooth function 
of $\epsilon$. If $|\varepsilon| \le \tau^q/q^2$ the function 
$\eta \sim 1$, beyond
this region it decreases as $|\varepsilon|$ increases.
As the barrier becomes more transparent, $\tau \rightarrow 1$, the
inter-peak distance (determined by the maximum value of
 $q$) decreases. Finally,
the peaks overlap forming the ballistic-like conductance $g =
\Delta /\hbar\nu$. 

To understand the result conjectured above let us consider the 
electron energy levels 
in the vicinity of
the Fermi level, $E_F$. Here the energy dispersion can be considered
to be linear. In a ballistic ring, one has then two sets of adiabatic 
energies $E_l(\Phi)$
corresponding to 
clock- and counterclockwise motion (Fig.~1). The sets coincide and 
scattering from the barrier opens gaps for those flux values where 
$E_l = E_F + l\Delta/2, \ l=0, \pm 1, \pm 2, \ldots$. 
Consequently,
the energy pumping into the system by a slowly varying magnetic flux
can be mapped to the one-dimensional motion of a quantum particle in
the field of periodically placed scatterers 
(cf. Refs.~\onlinecite{tg,gt,blatter}). 
Landau-Zener tunneling (with the amplitude
$\tau$ introduced above) through the gaps
corresponds to forward scattering while reflection from the
gaps is similar to backscattering. The important difference from the
usual impurity problem is that there is no translational invariance at
an arbitrary value of the driving force ${\cal E}$. This invariance is
only present for rational values $p/q$ of the dimensionless quantity
$A$\cite{blatter}. In this case we arrive at a super-lattice
containing  $q$ ``impurities'' per unit cell. As a result, the
motion along the $E$-axis is described by $q$ allowed bands, the
``velocity''  being $v_E \equiv {\dot E} \sim \Delta \tau^q/t_0$
(here $t_0 \equiv h/e{\cal E}$ is twice the time interval between
two sequential Landau-Zener scattering events). 
Since the upper bound of the Brillouin
zone is  $4\pi\hbar/q\Delta$,
the corresponding bandwidth for motion along the $E$-axis is $W \simeq
v_E4\pi\hbar/q\Delta \simeq 4\pi\hbar\tau^q/t_0 q$. 
At rational values $p/q$ of the quantity $A$ the electron rotates around 
the ring an
integer number, $2p$, of times during the time the enclosed magnetic flux 
changes by $q$ flux quanta. As a result motion along 
the $E$-axis can be mapped onto motion in a one-dimensional periodic 
potential, the corresponding eigenstates being extended. If $p/q$ is
irrational the motion along the $E$-axis is equivalent to motion in
a one-dimensional quasi-periodic potential. It turns out that the
corresponding states are then localized (see below) in spite of the fact that
there is no real disorder in the system. The localization radius,
$R_{loc}$, can be estimated for $A=p/q+\varepsilon,\ |\varepsilon| \ll 1/q$ 
as follows: In this
case the phase mismatch due to finite $\varepsilon$ can be thought of
as being due to a quasicassical potential $U(E)= \varepsilon \alpha E$
with $\alpha =8\pi \hbar/\Delta t_0$.
One
can see that a finite $\varepsilon$ gives rise to band bending which
creates semiclassical turning points for the propagating modes along the
$E$-axis. The localization length is in fact one half of the distance
between the turning points produced by the upper and lower band edges, 
$R_{loc} \simeq W/2\alpha |\varepsilon|=\Delta \tau^q/4q|\varepsilon|$.
Consequently, the localization time is
$t_{\text{loc}} \sim 4R_{loc}/v_E \sim t_0/q|\epsilon|$.

The manifestation of localization in the energy pumping depends on
the product $\nu t_{\text{loc}}$. When  $\nu t_{\text{loc}} \gg 1$ 
localization has not developed and the band picture of energy pumping
is relevant. The conductance is estimated as (cf. Ref.~\onlinecite{gt}) $G
=P/{\cal E}^2$, where $P$ is the average energy accumulation rate. The
quantity $P$, in its turn, is determined as $\nu \delta E\cdot N(\delta
E)$. Here $\delta E \sim v_E/\nu$ is the energy accumulated by a
single state, while $N(\delta E) \sim \delta E/\Delta$ is the number of
involved states. It follows that $g \sim \tau^{2q}
(\Delta/\hbar\nu)$. 
If $\nu t_{\text{loc}} \ll 1$, on the other hand,  $G$ is determined by hops between intra-band
localized states. In this case, $\delta E \sim 2R_{loc}$, and we obtain $g
\sim \hbar\nu \Delta \tau^{2q} /(e{\cal E}q\varepsilon)^2$. These
estimates are consistent with the first term in Eq.~(\ref{1}).

The consideration above are valid if the localization length is
greater than the unit cell size, $2R_{loc} \gg q\Delta/2$. In the opposite
limit the band picture fails and the velocity is dominated by
inter-band transitions. 
In this case, $\delta E \sim q\Delta, \ N(\delta
E) \sim q$, and $g \sim q^2 \hbar\nu \Delta/(e {\cal
E})^2$. This estimate corresponds to the second term in Eq.~(\ref{1}).

The following model is employed. The electron system is described by
the Hamiltonian 
\begin{equation} \label{2}
{\cal H} (t,\varphi) =- \Delta \sigma_z \left(
i\frac{\partial}{\partial \varphi} + \frac{t}{t_0} \right) +{\cal
H}_{\text{imp}} (\varphi), \ t_0 \equiv \frac{ch}{|e|{\dot \Phi}}
\end{equation}
Here $\sigma_i$ are Pauli matrices.
We are interested in the current, averaged over the time $t_0$,
\begin{equation} \label{3}
J(t)=\frac{1}{t_0}\int_{t-t_0/2}^{t+t_0/2} dt' \, \text{Tr} \, {\hat
\rho}{\hat \jmath}\, \quad {\hat \jmath} \equiv e {\dot{\hat {\varphi}}}\, .
\end{equation}
The single-electron density matrix, $\hat \rho$, is calculated from
the equation 
\begin{equation} \label{4}
\frac{\partial \hat\rho}{\partial t}= \frac{i}{\hbar}[{\hat \rho}, 
{\cal H}] - \nu
\left({\hat \rho}- f_0\bigl({\cal H}(t)\bigr)\right),
\end{equation}
where $f_0$ is the Fermi function. The formal solution of
Eq.~(\ref{4}) can be expressed in terms of the evolution operator
${\hat u} (t,t')$ for the systems described by the Hamiltonian
(\ref{2}),
\begin{equation} \label {5}
{\hat \rho} (t) =  \nu \int_{-\infty}^{t} dt' \, e^{\nu
(t'-t)}{\hat u}(t,t')  f_0  \bigl({\cal H} (t^\prime)\bigr){\hat
u}^{\dag}(t,t')\, .
\end{equation}
We are interested in the case of weak scattering, i.e. 
when the matrix element $V$ of the scattering potential corresponding
to a momentum transfer of $\sim 2p_F$ is much smaller than the inter-level
spacing $\Delta$. In such a situation, the impurity potential is
important only in the vicinity of the times when the adiabatic levels
for clockwise and counterclockwise motion cross. It creates  gaps
near the crossing points. Consequently, one can discriminate between
intervals of ballistic evolution (duration $\sim t_0/2$)  and
intervals of Landau-Zener 
scattering. The typical duration of such an interval is $\le \sqrt{t_0
\hbar/\Delta}$ (cf. Ref.~\onlinecite{lz}). Consequently, at $\Delta t_0/\hbar
\gg 1$ the Landau-Zener scattering is strongly confined within 
narrow intervals and can be described in terms of the scattering
matrix
\begin{equation} \label{s}
{\hat S}=\exp(i\theta)\left[ \rho\exp(i\xi \sigma_z)+i \tau
\sigma_x \right],  \ \rho = \sqrt{1-\tau^2}
\end{equation}
 where $\tau$ has been introduced above. It turns out that 
the physical values of interest here do not depend on
the phases $\theta$
and $\xi$. For simplucity we put them equal to zero.
The expression for $J$ has
 the form
\begin{equation} \label {J}
\frac{J}{J_0}=-\tau^2+\sum_{m=0}^\infty
e^{-2{\tilde \nu}m}\text{Tr} \,  {\hat T}^{-m} {\hat
J}{\hat T}^m {\hat {\cal F}}\, ,
\end{equation}
where $J_0=e\Delta/\hbar$ is the amplitude of the persistent current,
${\tilde \nu} \equiv \nu t_0/2$, \ ${\hat T}=\exp(-i{\hat
\varphi}){\hat u}(t_0,0)$, 
\begin{eqnarray} 
{\hat J}
&=&
\frac{1}{t_0} \int_{-t_0/2}^{t_0/2} dt\, {\hat u}^+(t,0) \sigma_z{\hat
u}(t,0) \, ,
\nonumber \\
{\hat{\cal F}}&=&\int_{-t_0/2}^{t_0/2} dt\,  {\hat u}^+(t,0)
 \frac{\partial f_0\bigl({\cal H}(t)\bigr)}{\partial t}{\hat
u}(t,0) \, . \nonumber
\end{eqnarray}
Here we have employed the relation ${\hat 
u}(t+mt_0,t'+mt_0)=\exp(im{\hat \varphi}){\hat 
u}(t,t')\exp(-im{\hat \varphi})$ which immediately follows from 
a similar symmetry
property of ${\hat {\cal H}}$. 

The problem of calculating the current  is reduced to an analysis
of the unitary 
operator $\hat T$. Having in mind the periodicity in $\varphi$ of all 
quantities we introduce the basis
$$
|n,s=\pm\rangle \equiv\frac{e^{\pm i(N_F+n)\varphi}}{\sqrt{2\pi}}{\bf s_\pm},
\quad {\bf s_+} =
\left(\begin{array}{c} 
1  \\ 0 \end{array} \right),
\
{\bf s_-} = \left(\begin{array}{c}
0 \\ 1 \end{array}\right) \, . 
$$
In this representation, the operator $\hat T$ can be expressed as a
direct product of operators acting in $n$ and pseudo-spin ($s$)
spaces,
$$ 
{\hat T}=
\rho{\hat S} e^{i\pi a\sigma_z}\otimes
e^{4i\pi a{\hat n}} 
+ i\tau \sum_\pm {\hat S}\sigma_\pm \otimes e^{2i\pi a{\hat n}}{\hat
R}_\pm e^{2i\pi a{\hat n}}\, ,
$$ 
where the operators ${\hat R}_\pm$ and $\hat n$ are defined as
$
{\hat R}_\pm |n,s\rangle=|n \mp 1, s \rangle$, and
${\hat n}
|n,s\rangle=n |n, s \rangle, 
$
and $a$ is the fractional part of the quantity $A$ introduced above,
$\sigma_\pm = (\sigma_x \pm  i \sigma_y)/2$.
In the $n,s$-representation
the operators ${\hat J}$ and ${\hat{\cal F}}$ have the form 
${\hat J}={\hat V} \delta_{n,n'}$, ${\hat{\cal F}}={\hat V}
\delta_{0,n}\delta_{0,n'}$, ${\hat V}= \tau^2\sigma_z -\rho \tau
\sigma_yf.$

It is straightforward to show that the unitary operator  
$\hat T$ has the following
properties:
$$
\hat R_- \hat T \hat R_+ = e^{-4i\pi a} \hat T ,  \, \,  
\sigma_y {\hat T}^* (-\hat n) \sigma_y = \hat T (\hat n) 
 $$
These result in the following relations between 
the eigenstates $\bbox{\psi}_\beta (n)$ of the operator $\hat T$
having eigenvalues 
$\exp(i\beta)$: 
\begin{eqnarray} \label{sp1}
\bbox{\psi}_\beta (n+1)& =& \bbox{\psi}_{\beta -4\pi a} (n)\, ,
\nonumber \\
\sigma_y \bbox{\psi}_\beta^* (-n)^& =&\bbox{\psi}_{-\beta} (n)\, .
\end{eqnarray}
Hence, the spectrum of $\hat T$ can be expressed in the form
$\exp(i\beta^\pm_r)$ where
$\beta^{\pm}_r=\pm \beta_0 (a) - 4\pi ar, \ r=0,\pm 1,\pm 2, \ldots$
One can prove that the properties (\ref{sp1}) allow one to
generate {\em a complete set } of eigenstates provided 
$\bbox{\psi}_{\beta_0}$ is known.
Furthermore, the
vector equation for the eigenstates of $\hat T$ reduces to a scalar equation
\begin{eqnarray}
B(m+1) &+& B(m-1)
\nonumber \\
 &&+\frac{2e^{i\pi a/2}}{\tau}\sin\left[\pi
a\left(m-\frac{1}{2}\right)+\frac{\beta}{2}\right] B(m) =0,
\nonumber
\end{eqnarray}
where $B$ is a linear combination of the components
$\bbox{\psi}_{\beta} $. Its solution allows one to determine
both the spectrum and the wave functions. The detailed analysis will be
published elsewhere\cite{we2}. 

The results are different for the cases of rational and
irrational values of $a$. 

\paragraph{Irrational $a$.} The analysis shows that $\beta_0 =\pi a$
and the set of $\{r\}$ is infinite. The eigenfunctions have the form
\begin{eqnarray} \label{ef}
\bbox{\psi}_{\pi a}
 &=&\frac{e^{\mp i \pi a n(2n\pm 1)}}{2\pi}
\int_0^{2\pi}dk\, e^{-2ikn}\left(\begin{array}{l}
e^{i\chi(k)}\\
e^{-i\chi(k-\pi a) +ik} \end{array}\right) ,
\nonumber \\ 
\chi(k)& =& -\sum_{l=1}^\infty \frac{\tau^l}{l} \, \frac{\cos
l(k-\pi/2)}{\sin \pi a l} \, .
\end{eqnarray}
It can be shown that the infinite series (\ref{ef}) converges
for almost all irrational values of $a$, and $\chi (k)$ is an
analytic function\cite{kassel}.  Consequently, the eigenfunctions
are exponentially localized, the localization length $R_{\text{loc}}$
in energy space  being
$$\frac{R_{\text{loc}}^2}{\Delta^2} 
\equiv \langle \beta|{\hat n}^2 |\beta \rangle - \langle
 \beta|{\hat n} |\beta \rangle^2 =
\sum_{l=1}^\infty\frac{\tau^{2l}}{4\sin^2 \pi a l}\, .$$ 
Here $|\beta \rangle \equiv
 \bbox{\psi}_{\beta}(n)$ is the eigenstate of the operator $\hat T$.
One can see that in the vicinity of rational values $p/q$ of $a$ the
localization length 
$R_{\text{loc}}$ diverges as $\Delta \tau^q/2\pi q|a-p/q|$ in agreement with
the qualitative estimates given above. 
The
expression for the dimensionless conductance $g$ at $\nu t_0 \ll 1$ yields
$$
g= \pi h \nu \frac{\Delta}{(e{\cal E})^2}\\
\left(\frac{2R_{loc}}{\Delta}\right)^2 
$$

\paragraph{Rational $a(=p/q)$.} Since the problem is translationally 
invariant in $n$-space. Consequently, the eigenstates can be labeled by a quasi
momentum ${\cal K}$. The spectrum is now given by\cite{b1}
\begin{equation}
\beta_0 ({\cal K})=\frac{\pi p}{q} - \frac{2}{q}\arcsin \left\{\tau^q
\sin \frac {q{\cal K}}{2}\right\} \, .
\end{equation}
 While the current can be expressed as
expression
\begin{eqnarray} \label{cur2}
\frac{J}{J_0}&=&\frac{1-\sqrt{1-\tau^{2q}}}{\tilde \nu} +
\sum_{\pm,r=0}^{q-1} \int d{\cal K}\, |\Omega^\pm_r|^2 \frac{{\tilde
\nu} }{{\tilde \nu}^2 + \sin^2 \Phi^\pm_r}\, ,
\nonumber \\
\Phi^\pm_r &=&\frac{\beta_0^+({\cal K})-\beta_r^\pm ({\cal
K})}{2},
\quad  \Omega^\pm_{r{\cal K}} = \sum_{l=0}^{q-1}
(\bbox{\phi}_{\beta_0},{\hat V}
\bbox{\phi}_{\beta_r^\pm})\, .
\end{eqnarray}
Here $\bbox{\phi}_{\beta_r^\pm}(l)$ is the Bloch amplitude
corresponding to the 
eigenstate $\psi_{r{\cal K}}^\pm (l)$.
An exact expression for the eigenfunction $\bbox{\psi}_{\beta_0}(n)$
shows the limiting transition to the expression (\ref{ef}) as $q,p
\rightarrow \infty, \ p/q = \text{const}$.
The current calculated according from Eq.~(\ref{cur2}) 
remains continuous also. Thus, Eq.~(\ref{cur2}) with large enough $q$ and $p$
can be used as a good approximation for irrational $a$-values. The results of
such a calculation is shown in Fig.~2.


The following two assumptions have been implicitly made in our 
discussion: (i) the electron
dynamics is governed by a linear dispersion law; 
(ii) the energy gaps as well as the scattering matrix $\hat S$
do not depend on the particular 
energy levels involved. Assumption (i) is
valid if the number of involved states (limited by the relaxation rate)
is much less than $N_F$. Estimates show that (i) is valid if
$(\hbar \nu/\Delta)^2 (\Delta/e{\cal E})N_F \gg 1$. The first factor in this
product is small while the two othes are large; the criterion
can be met under realistic experimental conditions. Assumption
(ii) is valid if the Fourier component of the barrier potential,
$\int V(\varphi)e^{2in\varphi}\, d\varphi$, is essentially
$n$-independent for relevant $n$. This is the case if the scattering
potential is confined to a region of width $\delta \varphi \ll
2\pi \nu t_0$. Note that the inequality $\nu t_0\ll 1$ is
essential for maintaining a noticeable energy pumping. 

Another approximation is that we have allowed 
for relaxation in the simplest possible way
by using a single relaxation time [cf. Eq.~(\ref{4})].
This is adequate if relaxation is caused by real space 
transfer of electrons between the ring and a surrounding reservoir.
If the electron energy level spectrum in the reservoir is continuous
the lifetime of an electron state in the ring with respect to this mechanism
is almost independent of quantum number. This mechanism allows us to
describe electron states in the ring as pure quantum states with a 
relaxation time given by the time of decay through escape to 
the reservoir. The exact results obtained above are relevant for the
case when such an ``escape" mechanism dominates. Internal inelastic relaxation
processes in the ring can lead to a significant difference between phase- and 
energy relaxation rates and requires a separate treatment. 
However, in the most interesting case of efficient Landau-Zener tunneling,
the intrinsic inelastic processes must involve large momentum transfers
and are therefore strongly suppressed \cite{ourprl}.

Finally, it would be interesting to discuss the cross-over from the 
above picture to a case where more than one scatterer introduces disorder
to the problem \cite{tg}.  Interference between reflections from different 
scatterers would induce an energy dependence of the gaps in the spectrum
and brake translational invariance in time. This would strongly affect the
quasi-ballistic regime and presumably suppress the absorption peaks in 
Fig.~2.

In conclusion the quantum electron dynamics problem in a
single-channel ballistic ring with a barrier subjected to a linearly
time-dependent magnetic flux has ben solved exactly. Exponential
localization in energy space has been proven. Finally, we have shown that the
dc-current has a set of peaks with fractional structure when plotted as a 
function of the induced electro-motive
force. This structure is strongly sensitive to the barrier height, as well 
as to the relaxation rate.

This work was supported by KVA, TFR and NFR. 
We also acknowledge partial financial support from INTAS grant N 94-3862.

\begin{figure}
\centerline{\psfig{figure=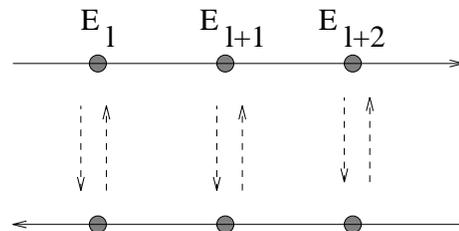,width=6cm}}
\caption{ Diagram showing coincidence of flux-driven energy levels
(corresponding to clockwise and anti-clockwise motion of electrons around ring)
at a special flux value (cf. text).}
\end{figure}

\begin{figure}
\centerline{\vspace*{-1in} \hspace*{-1in}\psfig{figure=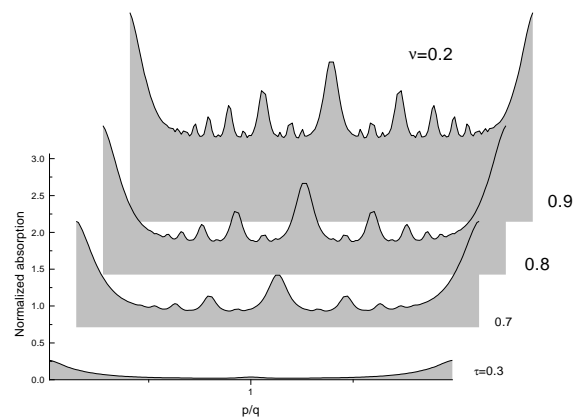,width=6cm}}
\caption{The normalized current $J/J_0$ as a function of
$a=p/q$ for different Landau-Zener tunneling amplitudes,
$\tau$. $J_0=e\Delta/\hbar$. A of ${\tilde \nu}=0.2$ for the
dimensionless relaxation rate was used. }
\end{figure}  


\begin{references}
\bibitem{b1} This expression was derived for $q=1$ in\cite{blatter}.
\end{references}
\end{document}